\documentclass[twocolumn,aps,prd,showpacs]{revtex4}
\usepackage{amssymb}

\newcommand{\sect}[1]{Sec.\,#1}
\newcommand{\Acknow}[1]{\subsection*{Acknowledgment} #1}
\def\noi{\noindent}
\def\nq{\hspace*{-1em}}

\def\nhq{\hspace*{-0.5em}}
\def\nhh{\hspace*{-0.3em}}
\def\cm{\hspace*{1cm}}

\def\Jl#1#2{{\it #1\/} {\bf #2},\ }

\def\ApJ#1 {\Jl{Astroph. J.}{#1}}
\def\CQG#1 {\Jl{Class. Qu. Grav.}{#1}}
\def\DAN#1 {\Jl{Dokl. AN SSSR}{#1}}
\def\GC#1 {\Jl{Grav. \& Cosmol.}{#1}}
\def\GRG#1 {\Jl{Gen. Rel. Grav.}{#1}}
\def\JETF#1 {\Jl{Zh. Eksp. Teor. Fiz.}{#1}}
\def\JETP#1 {\Jl{Sov. Phys. JETP}{#1}}
\def\JHEP#1 {\Jl{JHEP}{#1}}
\def\JMP#1 {\Jl{J. Math. Phys.}{#1}}
\def\NPB#1 {\Jl{Nucl. Phys.}{B\ #1}}
\def\NP#1 {\Jl{Nucl. Phys.}{#1}}
\def\PLA#1 {\Jl{Phys. Lett.}{#1A}}
\def\PLB#1 {\Jl{Phys. Lett.}{#1B}}
\def\PRD#1 {\Jl{Phys. Rev.}{D\ #1}}
\def\PRL#1 {\Jl{Phys. Rev. Lett.}{#1}}

\def\al{&\nhh}
\def\lal{&&\nhq{}}
\def\eq{Eq.\,}

\def\beq{\begin{equation}}
\def\eeq{\end{equation}}
\def\bear{\begin{eqnarray}}
\def\bearr{\begin{eqnarray} \lal}
\def\ear{\end{eqnarray}}
\def\earn{\nonumber \end{eqnarray}}
\def\nn{\nonumber\\ {}}

\def\nnn{\nonumber\\ \lal }

\def\eql{\al =\al}

\def\dst{\displaystyle}
\def\tst{\textstyle}
\def\fracd#1#2{{\dst\frac{#1}{#2}}}
\def\fract#1#2{{\tst\frac{#1}{#2}}}

\def\half{\fract 12}

\def\trih{\fract 32}

\def\eqdef{\stackrel{\rm def}=}
\def\e{{\,\rm e}}
\def\d{\partial}

\def\diag{\mathop{\rm diag}\nolimits}

\def\const{{\rm const}}
\def\eps{\varepsilon}

\newcommand{\vars}[1]{\left\{\begin{array}{ll}#1\end{array}\right.}

\def\ssph{static, spherically symmetric}

\def\bh{black hole}
\def\bhs{black holes}
\def\wh{wormhole}
\def\whs{wormholes}
\def\asflat{asymptotically flat}

\def\Sch{Schwarzschild}

\def\mn{_{\mu\nu}}

\def\mN{_\mu^\nu}

\def\R{{\mathbb R}}

\def\prad{p_{\rm rad}}
\def\schd{\biggl(1-\frac{2m}{r}\biggr)}

\def\cK{{\cal K}}

\begin{document}

\title {Possible wormholes in a brane world}
\author{K.A. Bronnikov}
\affiliation
  {VNIIMS, 3-1 M. Ulyanovoy St., Moscow 117313, Russia;\\
  Institute of Gravitation and Cosmology, PFUR,
        6 Miklukho-Maklaya St., Moscow 117198, Russia}
\email {kb@rgs.mccme.ru}

\author{Sung-Won Kim}
\affiliation
  {Dept. of Science Education, Ewha Womans University, Seoul 120-750, Korea}
\email {sungwon@mm.ewha.ac.kr}

\begin{abstract}
 The condition $R=0$, where $R$ is the four-dimensional scalar curvature,
 is used for obtaining a large class (with an arbitrary function of $r$) of
 static, spherically symmetric Lorentzian wormhole solutions. The wormholes
 are globally regular and traversable, can have throats of arbitrary size
 and can be both symmetric and asymmetric. These solutions may be treated as
 possible wormhole solutions in a brane world since they satisfy the vacuum
 Einstein equations on the brane where effective stress-energy is induced by
 interaction with the bulk gravitational field. Some particular examples
 are discussed.
\end{abstract}

\maketitle


\section{Introduction}

    Lorentzian wormholes as smooth bridges between different universes,
    or topological handles between remote parts of a single universe,
    have gained much attention since Morris, Thorne and Yurtsever discussed
    the connection between \whs\ and time machines \cite{thorne}; see
    \cite{vis-book,vis-hoh97} for reviews. It is well known that a \wh\
    geometry can only appear as a solution to the Einstein equations if the
    stress-energy tensor (SET) of matter violates the null energy condition
    (NEC) at least in a neighborhood of the \wh\ throat \cite{hoh-vis}.

    Many versions of exotic matter, able to provide NEC violation and to
    support \whs, have been suggested. One class of such sources is
    represented by so-called ghost fields, i.e., fields with explicitly
    negative energy density, including scalar-tensor theories of gravity
    with an anomalous sign of the scalar field kinetic term in the
    Lagrangian [5--13]
    Another class of static \whs\ is obtained with nonminimally coupled
    scalar fields \cite{br73,bar-vis99,sush-kim02} as a result of conformal
    continuation \cite{vac4}. The latter means that a singularity occurring
    on a certain surface $S$ in the Einstein frame metric, is removed by a
    conformal mapping to the Jordan frame, and the solution is then
    continued beyond $S$. It has been shown \cite{vac4} that a \wh\ is a
    generic result of a conformal continuation if its sufficient conditions
    are satisfied. In all such cases, however, the two \wh\ mouths are
    located in regions with different signs of the effective gravitational
    constant. In other words, if one mouth is in a normal gravity region,
    the other is in an antigravity region \cite{vac4}. A related problem is
    the instability of such \whs\ caused by the field behavior near the
    transition surface $S$ \cite{stepan}.

    Wormhole solutions have also been obtained in specific versions of
    dilaton gravity \cite{graf02} and gravity with torsion \cite{anch97}.
    Another approach is to invoke quantum effects, considering \whs\ as
    semiclassical objects [20--23] 
    (see also references therein). In all such cases, NEC violation is
    probably only possible in extremely strong gravity regions, leading to
    throat radii close to the Planck length. Wormholes thus seem to be an
    integral part of the hypothetic space-time foam but their practicability
    at macroscopic scales still remains vague.

    In our view, a natural source of \wh\ geometry can be found in the
    framework of the rapidly developing ideas of brane worlds (\cite
    {ransum1, ransum2}, for reviews see \cite{brane-rev}), inspired by the
    progress in superstring and M-theory \cite{ho-wit}. By this concept, the
    observable world is a kind of domain wall in a multidimensional space
    (5-dimensional in the simplest case), with large or even infinite extra
    dimensions. The standard-model fields are confined on the brane while
    gravity propagates in the surrounding bulk. The gravitational field on
    the brane itself can be described, at least in models of the type of the
    second Randall-Sundrum model \cite{ransum2}, by the modified
    4-dimensional Einstein equations derived by Shiromizu, Maeda and Sasaki
    \cite{maeda99} from 5-dimensional gravity with the aid of the Gauss and
    Codazzi equations. In vacuum, when matter on the brane is absent and
    the 4-dimensional cosmological constant is zero (a natural assumption
    for scales much smaller than the size of the Universe), these equations
    reduce to
\beq
          G\mn = - E\mn,                                         \label{EE}
\eeq
    where $G\mn$ is the 4-dimensional Einstein tensor corresponding to the
    brane metric $g\mn$ while $E\mn$ is the projection of the 5-dimensional
    Weyl tensor onto the brane. The traceless tensor $E\mn$ connects gravity
    on the brane with the bulk geometry (and is sometimes called the tidal
    SET), so that the set of equations
    (\ref{EE}) is not closed.  Due to its geometric origin, $E\mn$ does not
    necessarily satisfy the energy conditions applicable to ordinary matter.
    Thus, examples are known \cite{voll02a} when negative energies on the
    brane are induced by gravitational waves or black strings in the bulk.
    Therefore, if the brane world concept is taken seriously, $E\mn$ can be
    the most natural ``matter'' supporting \whs.

    In this paper we study \ssph, \asflat\ \wh\ solutions to the equation
    $R=0$, where $R$ is the 4-dimensional scalar curvature. Since $E\mn$ has
    zero trace, $R=0$ is an immediate consequence of (\ref{EE}). $R=0$ is a
    single equation connecting two metric functions, $\gamma(r)$ and $f(r)$,
    and can be solved with respect to $f$ for arbitrary $\gamma$.  We show
    that almost any $\gamma(r)$ satisfying some minimal requirements
    (smoothness and compatibility with asymptotic flatness) gives rise to a
    family of \wh\ solutions with the throat radius as a free parameter.
    Both symmetric and asymmetric \whs\ are obtained. We consider some
    particular examples, and, in addition to new \wh\ metrics, reproduce the
    results of the recent studies where some solutions with $R=0$ (though
    under other motivations) were found 
    [31--34]. As a by-product, some \bh\ solutions and solutions with naked
    singularities are also obtained. In the brane world framework, there
    remains a nontrivial problem to be solved: to inscribe the intrinsic
    brane geometry of the above solutions into the full 5-dimensional
    picture. Our 4-dimensional \wh\ solution can have an arbitrary size of
    the \wh\ throat, but a restriction can quite probably appear from
    5-dimensional geometry. Meanwhile, the present class of 4-metrics with
    zero scalar curvature can be of interest by itself.

    The paper is organized as follows. In \sect 2 we solve the equation
    $R=0$ and formulate the conditions under which the solution describes a
    symmetric or asymmetric \wh; in \sect 3 we discuss a few particular
    examples; \sect 4 contains some observations and concluding remarks.

\section{$R=0$: the general solution}

    The general \ssph\ metric in 4 dimensions in the curvature coordinates
    has the form
\beq                                                          \label{ds}
     ds^2 = \e^{2\gamma(r)} dt^2 - \e^{2\alpha(r)} dr^2 -r^2 d\Omega^2
\eeq
    where $d\Omega^2 = d\theta^2 + \sin^2 \theta\,d \phi^2$ is the linear
    element on a unit sphere\footnote%
{The sign conventions are: the metric signature $(+{}-{}-{}-)$; the curvature
 tensor $R^{\sigma}_{\ \mu\rho\nu} = \d_\nu\Gamma^{\sigma}_{\mu\rho}-\ldots$,
 so that, e.g., the Ricci scalar $R > 0$ for de Sitter space-time;
 and the stress-energy tensor such that $T^t_t$ is the energy density. }.

    The metric (\ref{ds}) gives, according to (\ref{EE}), the following
    expressions for the components of the effective SET $E\mN$, namely, the
    energy density $\rho = E^t_t$, the radial pressure $\prad = - E^r_r$ and
    the lateral pressure $p_\bot = -E^\theta_\theta = -E^\phi_\phi$:
\bear
    -\rho \eql \frac{1}{r^2}(\e^{-2\alpha}-1)                   \label{E00}
            - \frac{2\alpha_r}{r}\e^{-2\alpha};
\\                                                              \label{E11}
    \prad \eql \frac{1}{r^2}(\e^{-2\alpha}-1)
            + \frac{2\gamma_r}{r}\e^{-2\alpha};
\\                                                              \label{E22}
    p_\bot \eql \e^{-2\alpha} \Bigl( \gamma_{rr}+ \gamma_r^2
        - \alpha_r\gamma_r + \frac{\gamma_r - \alpha_r}{r}\Bigr),
\ear
    where the subscript $r$ denotes $d/dr$. In case $R=0$ one evidently has
    $2p_\bot = \rho - \prad$.

    Let us also write down the Kretschmann scalar for the metric (\ref{ds}):
\bear
    \cK \eql R\mn^{\ \rho\sigma}R{\rho\sigma}^{\ \mu\nu}
            = 4K_1^2 + 8K_2^2 + 8K_3^2 + 4K_4^2,
\nn
    K_1 \eql \e^{-2\alpha}(\gamma_{rr}+ \gamma_r^2 -\alpha_r\gamma_r),
\cm
    K_2 = \e^{-2\alpha} \frac{\gamma_r}{r},
\nn
    K_3 \eql -\frac{1}{r} \e^{-2\alpha}\alpha_r,
\cm
    K_4 = \frac{1}{r^2} (1-\e^{-2\alpha}).                      \label{Kr}
\ear
    The finiteness of $\cK$ is a natural regularity criterion for the
    geometries to be discussed. Indeed, $\cK$ is a sum of squares of all
    components $R\mn^{\eps\sigma}$ of the Riemann tensor for the metric
    (\ref{ds}), therefore $\cK < \infty$ is necessary and sufficient for
    finiteness of all algebraic curvature invariants.

    The condition $R=0$ which follows from (\ref{EE}) can be written as a
    linear first-order equation with respect to $f(r) \eqdef r\e^{-2\alpha}$:
\beq
    f_r (2 +  r\gamma_r)                                      \label{master}
         + f (2 r\gamma_{rr} + 2r\gamma_r^2 + 3 \gamma_r) = 2,
\eeq
    Its general solution is
\beq                                                          \label{sol}
      f(r) = \frac{2\e^{-2\gamma + 3\Gamma}}{(2 + r\gamma_r)^2}
     \int
      (2 + r\gamma_r) \e^{2\gamma - 3\Gamma}\, dr
\eeq
    where
\beq                                                          \label{Gamma}
     \Gamma(r) = \int \frac{\gamma_r dr}{2 + r\gamma_r}.
\eeq
    Thus, choosing the form of $\gamma(r)$ arbitrarily, we obtain $f(r)$
    from (\ref{sol}), and, after fixing the integration constant, the
    metric is known completely at least in the region where $\e^{\gamma}$
    and $\e^{\alpha}$ are smooth and nonzero.

    Let us now make clear how to choose the function $\gamma(r)$ (the
    so-called redshift function) and the integration constant in \eq
    (\ref{sol}) in order to obtain a \wh\ solution. We note for
    reference purposes that in many papers devoted to \whs, beginning with
    \cite{thorne}, the function $\e^{2\alpha}$ is expressed as
    $[1 - b(r)/r]^{-1}$ where $b(r)$ is the so-called shape function.
    Our $f(r)$ is then equal to $r-b(r)$.

    The coordinate $r$, which proves to be convenient for solving
    \eq(\ref{master}), is not an admissible coordinate in the whole space
    for \wh\ solutions since in this case $r$ has at least one minimum, and
    the solution in terms of $r$ therefore splits into at least two
    branches. As an admissible coordinate one can take, e.g., the Gaussian
    coordinate $l$ (proper length along the radial direction) connected with
    $r$ by the relation $l = \int \e^{\alpha} dr$, and the metric is
    rewritten as
\beq
     ds^2 = \e^{2\gamma(l)} dt^2 - dl^2 - r^2(l) d\Omega^2.   \label{ds1}
\eeq

    We seek static, traversable, twice \asflat\ \wh\ solutions. So we
    require: (i) there should be two flat asymptotics: $l \in \R$;
    $r\approx |l| \to \infty$ and $\gamma = \const + O(r^{-1})$ as $l\to
    \pm \infty$; (ii) both functions $r (l) > 0 $ and $\gamma(l)$ should be
    smooth (at least $C^2$) in the whole range $l \in \R$. This guarantees
    the absence of curvature singularities and horizons (the latter
    correspond to $\gamma \to -\infty$ which is ruled out). This also means
    that $r(l)$ should have at least one regular minimum, $r_{\min} > 0$
    (throat), at some value of $l$. Moreover, returning to functions of $r$,
    we see that at a flat asymptotic $\e^{\alpha} \to 1$ and $f(r)\approx r$.

    Suppose, without loss of generality, that a minimum of $r(l)$,
    that is, a \wh\ throat, is located at $l=0$. Then $r(0) = r_0 > 0$,
    $r_l(0) =0$ and (generically) $r_{ll}(0) > 0$, where the subscript $l$
    denotes $d/dl$. Near $l=0$ one has $r -r_0 \sim l^2$, hence the metric
    function $\e^{2\alpha(r)}$ behaves as $(r-r_0)^{-1}$, and $f(r) =
    r\e^{-2\alpha} \sim r-r_0$. In other words, a simple zero of $f(r)$ is
    an indicator of a \wh\ throat provided $\gamma(r)$ is smooth and finite
    at the same $r$.

    On the other hand, the derivative $\gamma_l(0)$ may be zero (which is
    always the case if the \wh\ is symmetric with respect to the throat) or
    nonzero. If $\gamma_l(0) = 0$, we shall have $\gamma_r (r_0) < \infty$.
    If, on the contrary, $\gamma_l (0) \neq 0$, then near $r_0$ we have
    $\gamma_r \sim 1/|l| \sim \sqrt{r-r_0}$, so that
\beq                                                         \label{asym}
    \gamma(r) \approx \gamma(r_0) + k \sqrt{r-r_0}, \cm k > 0.
\eeq
    We cannot put $k < 0$ since then we would obtain the expression $2 +
    r\gamma_r$ ranging from 2 (at spatial infinity) to $-\infty$ at $r=r_0$,
    so that $2 + r\gamma_r$ would vanish at some $r > r_0$ causing a
    singularity in (\ref{sol}).

    We are now ready to single out a class of symmetric \wh\ metrics (W1)
    and a class of potentially asymmetric \wh\ metrics (W2) on the basis of
    the solution (\ref{sol}).

\medskip\noi
    {\bf W1.} Specify the function $\gamma(r)$, smooth in the range $r_0
    \leq r <\infty$, $r_0 >0$, in such a way that $\gamma(\infty) =0$,
    $\gamma_r (r_0) < \infty$, and $2 + r\gamma_r > 0$ in the whole range.
    Fix the integration constant in (\ref{sol}) by performing
    integration from $r_0$ to $r$. Then these $\gamma(r)$ and $f(r)$
    determine a \wh\ which has a throat at $r=r_0$ and is symmetric with
    respect to it.

\medskip
    Indeed, by construction, $f(r) \sim r-r_0$ near $r_0$. Introducing the
    new coordinate $x$ by the relation $r = r_0 +x^2$, we have
    $\e^{2\alpha}dr^2 \sim (r-r_0)^{-1}dr^2 = 4 dx^2$, which leads to a
    perfectly regular metric whose all coefficients are even functions
    of $x \in \R$. Both $x\to +\infty$ and $x\to -\infty$ are flat
    asymptotics.

    Each $\gamma(r)$ chosen as prescribed creates a family of symmetric
    \whs\ with zero scalar curvature. The family is parametrized by the
    throat radius $r_0$, taking arbitrary values in the range where
    $\gamma(r)$ is regular and $2+r\gamma_r >0$.

    Another procedure is applicable to functions $\gamma(r)$ behaving
    according to \eq (\ref{asym}).

\medskip\noi
    {\bf W2-a.} Specify the function $\gamma(r)$, smooth in the range $r_0
    \leq r <\infty$, $r_0 >0$, such that $\gamma(\infty) =0$, $2 + r\gamma_r
    > 0$ in the whole range, and \eq (\ref{asym}) holds near $r_0$. Then,
    for proper values of the integration constant in (\ref{sol}), the sphere
    $r=r_0$ is a \wh\ throat, and the solution is smoothly continued beyond
    it.

\medskip
    Indeed, the solution (\ref{sol}) may be rewritten as follows:
\beq                                                          \label{sol-C}
    f(r) = \frac{\e^{-2\gamma + 3\Gamma}}{(1+\half r\gamma_r)^2}
       \biggl[   \int_{r_0}^r
      (1 + \half r\gamma_r) \e^{2\gamma - 3\Gamma}\, dr  + C\biggr]
\eeq
    Suppose $C > 0$. Then $f(r)$ behaves near $r_0$ as $r-r_0 =: x^2$, while
    $\gamma = \gamma(r_0) + kx + O(x^2)$.  The metric smoothly behaves at
    $r=r_0$ ($x=0$) in terms of the new coordinate $x$ and can be continued
    through this sphere. One cannot, however, guarantee that this
    continuation will lead to another flat spatial infinity to yield an
    asymmetic \wh, since the further behavior of $\gamma(x)$ and $f(x)$ may
    lead to a horizon or to a singularity.

    If we choose $C\leq 0$ in (\ref{sol-C}), we obtain two other situations:

\medskip\noi
    {\bf W2-b.} If $C < 0$, then $f(r_0) <0$; recalling that $f\sim r$ at
    large $r$, we see that $f(r)=0$ at some value $r = r_1 > r_0$, where
    $\gamma_r$ is finite, and we return to the circumstances described as
    W1, obtaining a symmetric \wh\ with $r\geq r_1$, and the sphere
    $r=r_1$ is its throat.

\medskip\noi
    {\bf W2-c.} If $C=0$, then near $r_0$ we obtain $f(r)\sim
    (r-r_0)^{3/2}$, and the metric is regularized at $r=r_0$ by another
    substitution: $r-r_0 = \xi^4$. As a result, \eq (\ref{asym}) yields
\[
    \gamma = \gamma(r_0) + k\xi^2 + \mbox{further even powers of}\ \xi,
\]
    and we again obtain a symmetric \wh, but now with a quartic behavior of
    $r$ near its minimum as a function of the admissible coordinate $\xi
    \in\R$.

\section{Examples}

    We will present expressions for the metric functions $\gamma$ and $f$,
    the effective ``tidal'' energy density $\rho$ and the sum $\rho +
    \prad$, which characterizes violation of the null energy condition (for
    \ssph\ systems this condition reduces to $\rho + \prad \geq 0$).

    We use the time scale of a remote observer at rest and so
    always assume that $\e^\gamma \to 1$ as $r\to \infty$.

\medskip\noi
    {\bf 1.} The simplest example is obtained for $\gamma\equiv 0$.
    Choosing any $r_0>0$ and applying the W1 algorithm of \sect 2, we simply
    obtain $f(r) = r-r_0$.  This is a symmetric \wh\ solution known as the
    spatial \Sch\ geometry \cite{dadh}:
\bear                                                            \label{wh1}
    ds^2 \eql dt^2 - \biggl(1-\frac{r}{r_0}\biggr)^{-1}dr^2 -r^2 d\Omega^2
\nn
     \eql dt^2 - 4(r_0 + x^2) dx^2 - (r_0 + x^2)^2 d\Omega^2.
\ear
    The effective SET $E\mN$ has the form
    $T\mN = \diag (0,\ -p_r,\ p_r/2,\ p_r/2)$ with the radial pressure
\beq
       p_r = -r_0/r^3.                                           \label{p1}
\eeq

\medskip\noi
    {\bf 2.} Our next example uses the \Sch\ form of $\gamma$:
\beq
     \e^{2\gamma} = 1- \frac{2m}{r},\cm  m > 0.              \label{ga2}
\eeq
    Choosing any $r_0 > 2m$, we obtain according to the W1 prescription:
\bear
    f(r) \eql \frac{(r-2m)(r-r_0)}{r-3m/2},                  \label{f2}
\\
    ds^2 \eql \schd dt^2
    	- \frac{\Bigl(1 - \fracd{3m}{2r}\Bigr) dr^2 }
	{\Bigl(1 - \fracd{2m}{r}\Bigr)\Bigl(1 - \fracd{r_0}{r}\Bigr)}
        -r^2 d\Omega^2
\nn                                                           \label{wh2}
     \eql \frac{x^2 + r_0 - 2m}{r_0+x^2} dt^2
\nnn \hspace{-15mm}
          - \frac{4 (r_0+x^2) (r_0 +x^2 - \trih m)}{x^2 + r_0 -2m} dx^2
          - (x^2 + r_0)^2 d\Omega^2.
\ear
    This is evidently a symmetric \wh\ geometry for any $r_0 > 2m \geq 0$,
    or for any $r_0 >0$ in case $m < 0$. The \Sch\ metric is restored
    from (\ref{wh2}) in the special case $r_0 = 3m/2$.

    The SET components of interest are
\bear
     \rho \eql \frac{m(r_0- \trih m)}{2r^2(r - \trih m)^2};    \label{SET2}
\nn
     \rho + \prad \eql - \frac{(r-2m)(r_0 - \trih m)}{r^2 (r-\trih m)^2}.
\ear

    The metric (\ref{wh2}) was obtained by Casadio, Fabbri and Mazzacurati
    \cite{casad01} in search for new brane-world \bhs\ and by Germani and
    Maartens \cite{maart01} as a possible external metric of a homogeneous
    star on the brane, but the existence of traversable \wh\ solutions for
    $r_0 > 2m$ (in the present notations) was not mentioned. It was supposed
    in \cite{casad01} that the post-Newtonian parameters of the metric must
    be close to their Einstein values for experimental reasons and therefore
    restricted their study to configurations close to \Sch. Then $r_0$
    must be close to $3m/2$. In this case, as in the \Sch\ metric, $r=2m$ is
    an event horizon, but, according to \cite{casad01}, the space-time
    structure depends on the sign of $\eta = r_0 - 3m/2$. If $\eta < 0$, the
    structure is that of a \Sch\ \bh, but the curvature singularity is
    located at $r= 3m/2$ instead of $r=0$. If $\eta >0$, the solution
    describes a nonsingular \bh\ with a \wh\ throat at $r=r_0$ inside the
    horizon, in other words, a non-traversable \wh\ \cite{casad01}.

    We would here remark that, in our view, such hypothetic objects as
    brane-world \bhs\ or \whs, not necessarily of astrophysical size,
    need not necessarily conform to the restrictions on the post-Newtonian
    parameters obtained from the Solar system and binary pulsar
    observations, and it therefore makes sense to discuss the full range of
    parameters which are present in the solutions.

\medskip\noi
    {\bf 3.} Consider the extreme Reissner-Nordstr\"om form of $\gamma(r)$:
\beq
     \e^{2\gamma} = \schd^2, \cm m > 0.                        \label{ga3}
\eeq
    The W1 procedure now leads to
\beq                                         		\label{f3}
    f(r) = \frac{(r-r_0)(r-r_1)}{r},
            \cm       r_1 \eqdef \frac{mr_0}{r_0-m};
\eeq
\bearr
    ds^2 = \schd^2 dt^2 -                                    \label{wh3}
	 \frac{r^2\, dr^2}{(r-r_0)(r-r_1)} - r^2 d\Omega^2
\nnn    \nq
     = \biggl(1 - \frac{2m}{r_0 {+} x^2}\biggr)^2 dt^2
         - 4 \frac{(r_0 {+} x^2)^2 dx^2}{r_0 - r_1 + x^2}
         - (r_0 {+} x^2) d\Omega^2.
\nnn
\ear
    where we assume $r_0 > 2m$, so that $r_1 < r_0$. This is a symmetric
    \wh\ metric. The SET components of interest are
\bear
     \rho \eql \frac{mr_0^2}{r^4(r_0-m)},                       \label{SET3}
\nn
     \rho + \prad \eql - \frac{(r_0-2m)^2}{r^2(r-2m)(r_0-m))}.
\ear

    In the solution (\ref{f3}), $r_0$ may be regarded as an integration
    constant, so it is of interest what happens if $r_0 \leq 2m$.
    Evidently, $r_0 = 2m$ leads to the extreme Reissner-Nordstr\"om
    \bh\ metric (which is well known to possess a zero Ricci scalar, as does
    the general Reissner-Nordstr\"om metric).
    In case $2m > r_0 > m$, we have $r_1 > 2m$, and we again obtain a
    symmetric \wh, but now $r$ ranges from $r_1$ to infinity and $r=r_1$ is
    the throat. Actually, $r_0$ and $r_1$ exchange their roles as compared
    with the case $r_0 > 2m$. This property was expected due to symmetry
    between $r_0$ and $r_1$ in the metric (\ref{wh3}).

    The value $r_0 = m$ is meaningless. Lastly, $r_0 < m$ leads either to
    $r_1 < 0$ (for $r_0 \geq 0$) or to $0 < r_1 < 2m$ (for $r_0 < 0$). The
    solution exists in both cases for $r > 2m$ only, and $r=2m$ turns out to
    be a naked singularity, as is confirmed by calculating the Kretschmann
    scalar.

\medskip\noi
    {\bf 4.} Consider an example belonging to class W2
    described in the previous section. Namely, let us choose
\bear                                                          \label{ga4}
    \e^{2\gamma} = \left(1 - b + b \sqrt{1 - 2m/r}\right)^2,
\ear
    with $b = \const\neq 0$. The special cases $b=0$ and $b=1$ nave been
    already discussed in Examples 1 and 2, respectively. The form (\ref{ga4})
    of $\e^{\gamma(r)}$ has been found \cite{dadh, camera02, casad01} by
    solving the equation $R=0$ under the condition that the energy density
    $T^0_0$ is zero, whence it followed that $\e^{-2\alpha} = 1-2m/r$, and,
    in our notation, $f(r) = r-2m$.  Note that the Schwarzschild mass, found
    from the large $r$ behavior of $\gamma(r)$, is equal to $bm$ rather
    than $m$.

    Knowing that $f(r)=r-2m$ is a special solution to the inhomogeneous
    equation (\ref{master}) with $\gamma(r)$ given by (\ref{ga4}), we can
    make easier the integration in (\ref{sol}) by writing the solution as
    $f(r) = 2r-m + f_1(r)$ where $f_1$ is a general solution to the
    corresponding homogeneous equation. We obtain
\bear                                                            \label{f4}
      f(r) = r - 2m + C\frac{\e^{-2\gamma + 3\Gamma}}{(2 + r\gamma_r)^2},
\ear
    where $C_1=\const$ and $\Gamma$ has been defined in (\ref{Gamma}). The
    form of $\Gamma(r)$ depends on the constant $c \eqdef 2(1-b)/b$
    (the case $b=0$ is excluded):
\bearr
    \e^{3\Gamma} =
\nnn
     \vars{                                       \label{Ga4}
      (1 + 2cv + 3v^2)
      \exp \biggl [-\fracd{2c}{c'}\arctan\fracd{c{+}3v}{c'}\biggr],
                            & c < \sqrt{3};
\\
      (1 + \sqrt{3}v)^2 \, \exp\Bigl[2/(1+\sqrt{3}v)\Bigr]
                            & c = \sqrt{3};
\\
      (1 + 2cv + 3v^2)\biggl[\fracd{c+3v+c'}{c+3v-c'}\biggr]^{-c/c'},
                            & c > \sqrt{3},   }
\nnn
\ear
    where we have denoted
\beq
     c' = \sqrt{|c^2-3|}, \cm v = \sqrt{1-\frac{2m}{r}}.      \label{def-v}
\eeq

    For all three cases in (\ref{Ga4}), depending on the integration
    constant $C_1$, one can single out the behaviors of classes W2-a,
    W2-b and W2-c (in their description in \sect 2 one should substitute
    $r_0=2m$).  The critical value of $C_1$, corresponding to $C=0$ in \eq
    (\ref{sol-C}), is
\beq
      C_{1\rm cr} = -2m \e^{-3\Gamma}\Bigr|_{v=0}.              \label{C-cr}
\eeq
    It corresponds to integration in \eq(\ref{sol}) from $r_0$ to $r$, and
    the solution then belongs to class W2-c, a symmetric \wh\ with a quartic
    dependence of $r$ on the admissible coordinate $\xi$.

    The case $C_1 < C_{1\rm cr}$ corresponds to integration in (\ref{sol})
    from some $C_1$-dependent radius $r_1 > 2m$ to $r$, and the solution
    belongs to type W2-b equivalent to W1: a symmetric \wh\ with an
    arbitrary throat radius $r_1 > 2m$. We will not write down the full
    cumbersome expressions for the metric for $C_1 \leq  C_{1\rm cr}$ since
    the qualitative properties of the solutions are already clear.

    Of greater interest are solutions with $C_1 > C_{1\rm cr}$, for which
    $r=2m$ is an asymmetric throat [class W2-a]. This is the only class in
    which the metric continued beyond the throat behaves ``individually'',
    i.e., depends on the specific choice of $\gamma(r)$ and $C_1$ rather
    than follows the above general description.

    A good coordinate for passing the throat $r=2m$, more convenient than
    the previously used coordinate $x$, is $v$ defined in (\ref{def-v}). We
    have $r(v) = 2m/(1-v^2)$; the original spatial asymptotic corresponds to
    $v = 1$, the throat is located at $v=0$, and another spatial asymptotic
    can be located at $v=-1$ if the metric avoids singularities on the way
    to it. Let us find out whether it is the case.

    The requirement $\e^{\gamma} > 0$ at $v\geq -1$ leads to $b < 1/2$,
    hence $c >2$, and we are left with the third line in the expression
    (\ref{Ga4}) for $\Gamma(v)$. Then the metric coefficient $g_{vv}$ is
    given by
\bearr        \nq
     - g_{vv} = \frac{16m^2}{(1-v^2)^2}                     \label{g_vv}
\nnn \times
        \biggr[1 + C_1 \frac{1-v^2}{6m}
            (v{-}v_-)^{3v_-/c'} (v{-}v_+)^{-3v_+/c'}\biggr]^{-1},
\ear
    where $v_\pm$ are roots of the trinomial $P(v) = 3v^2 + 2cv +1$:
\[
     v_{\pm} = \frac{1}{3} \left(-c \pm \sqrt{c^2-3}\right)
     		=\frac{1}{3}(-c\pm c').
\]
    For $c > 2$, we have $v_- < -1$ whereas the other root $v_+$ lies
    between 0 and -1. On the other hand, $-3v_+/c'$, i.e., the
    exponent of the binomial $(v-v_+)$ in (\ref{g_vv}), is a number between
    0 and 1. Therefore $g_{vv}$ is finite at $v=v_+$ but has an infinite
    derivative with respect to $v$. Transforming back to $r$ (for $v <0$,
    the transformation is $v = - \sqrt{1-2m/r}$), we observe that the metric
    coefficient $g_{rr} = -e^{2\alpha}$ [see (\ref{ds})] is singular at
    $r = r_+ = 2m/(1-v_+)^2$. More precisely, $\e^{2\alpha}$ is finite but
    contains a term proportional to $(r-r_+)^k$ where $0 < k< 1$, hence
    $\alpha_r \sim (r-r_+)^{k-1} \to \infty$, and the Kretschman scalar
    (\ref{Kr}) blows up due to the divergence of its constituents $K_1$ and
    $K_3$.

    We conclude that, under the choice (\ref{ga4}) of $\gamma(r)$, the class
    of solutions (W2-a) does not contain \wh\ solutions. Having passed the
    throat at $r=2m$ ($v=0$), we ultimately arrive at a singularity or maybe
    a horizon, which is not excluded in case $b > 1/2$.

    An exception is the case $C_1 = 0$, when we return to the solution known
    from Refs.\,\cite{dadh, camera02, casad01}, which has been described at
    length in these papers. We will only mention the main points in our
    notations. The metric in terms of $v$ is
\beq
     ds^2 = (1 -b +bv)^2 dt^2 - \frac{16\,m^2}{(1 - v^2)^4}dv^2
            - \frac{4m^2}{(1 - v^2)^2} d\Omega^2.            \label{ds4}
\eeq
    In case $b=1$ it is another form of the \Sch\ metric. For $b > 1/2$ but
    $b \neq 1$, the sphere $v = (b-1)/b$ is a naked singularity, as may be
    concluded from the fact that $\gamma_r \to \infty$ while $\alpha$ and
    $r$ are finite, hence the quantity $K_2$ in (\ref{Kr}) blows up. This
    singularity is located at positive $v$, i.e., before reaching the
    throat $v=0$, if $b>1$ and at negative $v$, beyond the throat, if $b<1$.
    Note that for $v < 0$ we have in the curvature coordinates
    $g_{tt} = \e^{2\gamma} = 1-b-b\sqrt{1-2m/r}$.

    In case $b < 1/2$ the metric (\ref{ds4}) describes an asymmetric \wh\
    even having different signs of mass at its two flat asymptotics: the
    mass is equal to $bm$ at $v=1$ and to $-bm/(1-2b)$ at $v=-1$.

    Lastly, if $b=1/2$, then $v=-1$ is a horizon having an infinite area and
    zero Hawking temperature, like the previously described cold \bhs\ in
    scalar-tensor theories of gravity \cite{cold}. The spatial part of the
    metric is flat at $v\to -1$. Moreover, as is directly verified, the
    canonical parameter for timelike, spacelike or null geodesics takes an
    infinite value at at $v=1$, which means that this space-time is
    geodesically complete (as are \wh\ space-times), and no further
    continuation is required.

\section {Concluding remarks}

    We have seen that the equation $R=0$ leads to a great number of \wh\
    solutions. Symmetric \wh\ solutions of class W1 can be obtained from
    any $\gamma(r)$ providing asymptotic flatness; asymmetric \wh\ solutions
    belonging to class W2-a require somewhat more special conditions. As
    follows from Examples 1-3, \whs\ are not always connected with negative
    (effective) energy densities $\rho$; they can appear with $\rho >0$, but
    only with comparatively large negative pressures maintaining violation of
    the null energy condition.  Example 3 shows that, for given $\gamma(r)$,
    sometimes even more \wh\ solutions can be obtained than was expected in
    search for class W1 solutions. Example 4 shows that asymmetric \whs\ are
    more difficult to obtain from the general solution (\ref{sol}) than
    symmetric ones.

    Black-hole solutions can also be obtained from (\ref{sol}) but under
    more restrictive conditions. Indeed, given a specific function
    $\e^{\gamma(r)}$ increasing from zero at some $r=r_h$ to 1 at $r=\infty$,
    a \wh\ solution to $R=0$ can be obtained with a throat at any $r>r_h$
    whereas in a \bh\ solution the event horizon is fixed at $r=r_h$.
    Positive functions $\e^{\gamma(r)}$ lead to numerous \wh\ solutions but
    not \bh\ ones. Thus, roughly speaking, \whs\ as solutions to $R=0$ are
    more numerous than \bhs.

    All this referred to metrics satisfying the condition $R=0$
    in 4 dimensions, which admits an interpretation as the brane metric.
    It has been claimed that ``any 4-dimensional space-time with $R=0$ gives
    rise to a 3-brane world without surface stresses embedded in a
    5-dimensional space-time'' \cite{voll02b} since the embedding
    contains a very significant arbitrariness. Nevertheless, a complete
    model requires knowledge of the full 5-dimensional space-time.
    In other words, one should ``evolve'' the 4-metric into the bulk, using
    this 4-matric as initial data for the 5-dimensional equations. It is
    rather a difficult task, as was demonstrated in a study of particular
    \bh\ solutions in Refs. \cite{chamb01,casad02}. There are, however, two
    favorable circumstances. One is the wealth of \wh\ solutions: there
    is actually an arbitrary function $\gamma(r)$ leading to \whs\ on the
    brane, which must in turn lead to a wide choice of suitable bulk
    functions. The other is the global regularity of \wh\ space-times, and
    one can expect that the bulk incorporating them will also be regular.
    (It may be recalled that it was the singular nature of \bh\ solutions
    that caused some technical difficulties in Ref.\,\cite{chamb01}.) We
    hope that it will be possible to obtain meaningful complete \wh\
    models within the brane world concept; the work is in progress.

\Acknow
{The work was supported in part by grant No. R01-2000-000-00015-0 from
the Korea Science and Engineering Foundation and in part by Asia
Pacific Center for Theoretical Physics. KB is grateful to the colleagues
from Ewha Womans University, Seoul, for kind hospitality during his stay in
October-November 2002.}

\end{document}